\documentclass[aps,pra,reprint,superscriptaddress]{revtex4-2}

\usepackage{amssymb}
\usepackage{amsmath}
\usepackage{dcolumn}
\usepackage{graphicx}
\usepackage{mathrsfs}
\usepackage{appendix}
\usepackage{graphicx}
\usepackage{booktabs}
\usepackage{colortbl}
\usepackage{float}

\definecolor{Dred}{RGB}{190,0,0}

\usepackage{url}
\usepackage[colorlinks]{hyperref}
\hypersetup{%
	plainpages=true,
	breaklinks=true,       
	hypertexnames=false,  
	pageanchor=true,
	colorlinks=true,
	linkcolor={blue},
	citecolor={red},
	urlcolor={blue},
	anchorcolor={black}
}

\def \hide#1{}

\begin{document}
\title{Synthesizing In-Bulk Topological Corner States via Giant Atoms}

\author{Zhao-Min Gao}
\affiliation{Shaanxi Province Key Laboratory of Quantum Information and Quantum Optoelectronic Devices, School of Physics, Xi'an Jiaotong University, Xi'an 710049, People’s Republic of China}

\author{Xin Wang}
\email{wangxin.phy@xjtu.edu.cn}
\affiliation{Shaanxi Province Key Laboratory of Quantum Information and Quantum Optoelectronic Devices, School of Physics, Xi'an Jiaotong University, Xi'an 710049, People’s Republic of China}

\date{\today}

\begin{abstract} 
Corner states in higher-order topological insulators are typically confined to geometric corners, limiting their flexibility for scalable quantum information processing. We propose a scheme to synthesize topological corner states at arbitrary positions within the bulk of a two-dimensional Su-Schrieffer-Heeger (SSH) lattice by coupling it to giant atoms. By engineering an L-shaped multi-point coupling that satisfies the vacancy-like dressed state (VDS) condition, where the photonic wavefunction vanishes at the coupling sites to form an artificial bulk boundary, we derive the conditions for synthesizing a zero-energy corner state at any target position. We demonstrate that the engineered corner state exhibits high fidelity and spatial localization, remaining robust against realistic disorder. Extending to multi-atom networks, we realize a versatile quantum switch via a giant superatom, enabling multi-channel control over 0D corner states and 1D edge states through the dual-resonance condition. Furthermore, we demonstrate the coherent interactions between two giant atoms mediated by VDS-engineered corner states. Governed by a sublattice selection rule, the coupling activates exclusively in intersecting configurations and decays exponentially with distance. Our work establishes a highly reconfigurable platform for embedding topological boundary modes within the bulk, offering a robust pathway for scalable topological quantum networks.

\end{abstract}
\maketitle
\section{Introduction}
Higher-order topological insulators extend the conventional bulk-boundary correspondence by hosting symmetry-protected topological corner states ~\cite{Wladimir_2017,PhysRevLett.119.246402,schindler2018higher,PhysRevB.98.205147,PhysRevLett.132.176302,PhysRevB.111.115418,PhysRevB.105.L081107,serra2018observation,PhysRevLett.125.056402,rlk2-psxm}. These zero-dimensional modes are governed by bulk topological invariants and exhibit strong robustness against disorder and fabrication variations, making them suitable as robust information carriers in photonic and quantum information platforms~\cite{PhysRevResearch.7.023079,noh2018topological,Yang_2024,PhysRevB.96.245115,PhysRevA.96.043811,PhysRevB.110.024402,PhysRevB.99.085406,PhysRevB.102.214204,070710-2026-0534,Xie:21,PhysRevLett.122.233903,Chen2026Acoustic}. However, these corner states are conventionally confined to the geometric corners of the lattice. This geometric confinement restricts their flexibility and limits their application in scalable topological quantum networks. Therefore, developing methods to synthesize corner states at arbitrary positions is essential for advancing flexible topological devices.

Several approaches have been explored to overcome this geometric constraint, including the introduction of local on-site potentials, generalized quadrupole models, and interlayer coupling in twisted bilayer systems~\cite{SUN2025667,PhysRevB.110.L121301,PhysRevB.111.165418}. While these pioneering works have demonstrated the tunability of corner-like modes through global parameter modulation or lattice restructuring, creating and manipulating topological states at arbitrary locations without altering the underlying bulk lattice remains a significant challenge. Achieving such local, on-demand reconfigurability is highly desirable for building flexible and scalable quantum networks.

Giant atoms, which couple nonlocally to multiple sites of a quantum lattice, offer a promising platform for engineering light-matter interactions through quantum interference effects~\cite{Nature_superconducting,PhysRevLett.120.140404,PhysRevA.111.023711,PhysRevA.106.033522,Bello_2019,PhysRevA.90.013837,Jia_24,PhysRevResearch.6.043222,PhysRevA.107.013710,PhysRevLett.126.043602,PhysRevLett.128.223602,PhysRevA.107.023705,PhysRevA.109.063708}. For atom-photon coupling, the vacancy-like dressed state (VDS) emerges as a distinct phenomenon where destructive interference forces the photonic wavefunction to exhibit zero amplitude at the coupling sites~\cite{PhysRevLett.126.063601,Leonforte_2025,B_nsel_2026,PhysRevA.107.043714,PhysRevB.107.054301}. By harnessing the VDS effect, the multi-point coupling geometry of a giant atom can be designed to generate artificial boundaries through the spatial arrangement of its coupling points. This capability provides a powerful route to induce effective topological boundaries at desired positions, decoupling topological localization from physical geometry.

In this work, we propose a platform where giant atoms interact with a two-dimensional Su-Schrieffer-Heeger (SSH) lattice, a conventional model supporting topological states ranging from 2D bulk bands and 1D edge modes to 0D corner states~\cite{Xie2021Higherorder,PhysRevA.101.063839,PhysRevB.100.075437,Wang_2024,PhysRevA.109.022211,37gn-2g16,Shi_2026,Peng_2026,PhysRevB.107.045118,https://doi.org/10.1002/lpor.202400638,Mandal_2026,PhysRevLett.122.233902}. By harnessing the VDS mechanism and engineering the giant atom’s multi-point coupling, we analytically derive the exact conditions required to create an artificial boundary, thereby synthesizing robust zero-energy corner states at arbitrary positions within the bulk. Furthermore, we extend this framework to multi-atom networks. We introduce a giant superatom configuration to realize a versatile quantum switch, enabling multi-channel control over corner states and edge states via dual-resonance engineering. Additionally, we demonstrate that two spatially separated giant atoms can interact coherently via their VDS-engineered corner states, exhibiting strong geometry-dependent coupling. By decoupling topological localization from physical boundaries, our scheme enables the on-demand creation of topological boundary modes within the bulk, opening a flexible avenue toward scalable topological quantum architectures.

\begin{figure}
	\centering
	\includegraphics[width=0.8\linewidth]{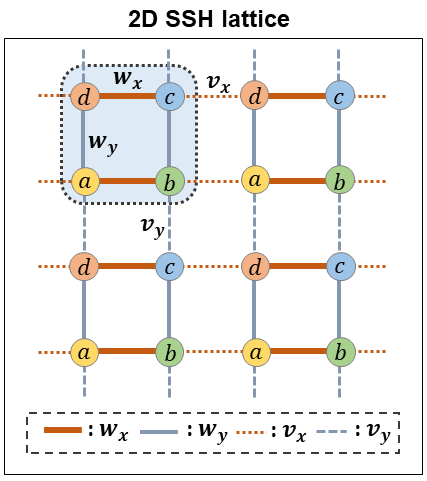}
	\caption{Schematic diagram of two-dimensional SSH model. Each unit cell (shaded region) contains four sublattices labeled $a, b, c,$ and $d$. Intracell (intercell) coupling strengths along x and y directions are denoted as $w_{x,y}$ ($v_{x,y}$), represented by solid (dashed) lines.}
	\label{fig1}
\end{figure}

\section{Higher-Order Topological Phases in the 2D SSH Model}
We consider a two-dimensional  SSH lattice, which is a typical platform for higher-order topological insulators. As illustrated in Fig.~\ref{fig1}, a unit cell consists of four sublattices, labeled $a, b, c,$ and $d$. The intracell coupling strengths along the $x$ and $y$ directions are denoted by $w_x$ and $w_y$, respectively, while the intercell coupling strengths between neighboring cells are given by $v_x$ and $v_y$~\cite{acta_ssh}. By tuning the ratio between intracell and intercell coupling strengths, the system undergoes a topological phase transition, hosting higher-order topological states such as edge and corner modes. The Hamiltonian of the 2D SSH model is written as
\begin{align}
	\!\!\!\! H_B \!=\!\!\sum_{i,j}\!{w_x \!\left(\! b_{i,j}^{\dagger}a_{i,j}+c_{i,j}^{\dagger}d_{i,j}\! \right)\!\!+\! v_x \!\left(\! a_{i+1,j}^{\dagger}b_{i,j}\!+\! d_{i+1,j}^{\dagger}c_{i,j} \!\right)} \notag\\
	\!\!+w_y\!\left(\! c_{i,j}^{\dagger}b_{i,j}+d_{i,j}^{\dagger}a_{i,j}\! \right)\!+\! v_y\!\left(\! b_{i,j+1}^{\dagger}c_{i,j}+a_{i,j+1}^{\dagger}d_{i,j}\! \right)\! +\! \text{H.c.},
	\label{H_B}
\end{align}
where $i$ and $j$ denote the unit cell indices along the $x$ and $y$ directions. We define the topological parameter $\beta=v_{x}/w_{x}=v_y/w_y$ to characterize the topological phases. When $\beta > 1$, the system enters a non-trivial higher-order topological insulator phase characterized by quantized bulk polarization and Wilson loop invariants, hosting symmetry-protected zero-energy corner states~\cite{PhysRevB.96.245115,PhysRevLett.132.213801,PhysRevA.106.063524}.

\begin{figure}[H]
	\centering
	\includegraphics[width=0.9\linewidth]{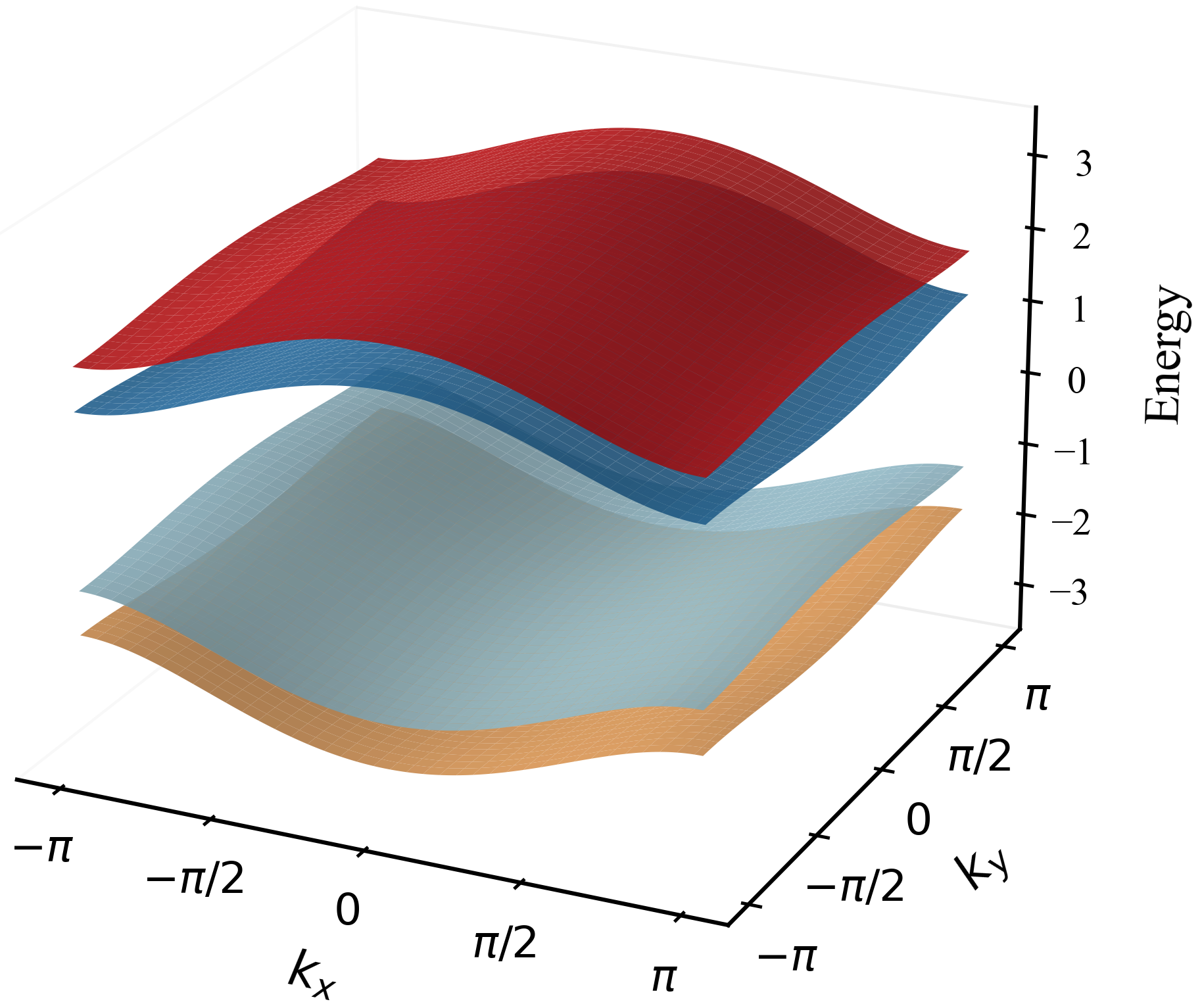}
	\caption{ Bulk band structure of the 2D SSH model under periodic boundary conditions (PBC). The parameters are set to $w_x=0.5$, $w_y=0.1$, $v_x=2.0$, and $v_y=0.4$.}
	\label{fig2}
\end{figure}

Under periodic boundary conditions (PBC), the bulk Hamiltonian is diagonalized in momentum space. In the Bloch basis $\Psi_{\mathbf{k}} = (a_{\mathbf{k}}, b_{\mathbf{k}}, c_{\mathbf{k}}, d_{\mathbf{k}})^T$, the system is described by $H_B = \sum_{\mathbf{k}} \Psi_{\mathbf{k}}^\dagger \mathcal{H}(\mathbf{k}) \Psi_{\mathbf{k}}$, where the $4\times4$ Bloch Hamiltonian $\mathcal{H}(\mathbf{k})$ reads~\cite{Ni_2017,PhysRevB.84.195452}
\begin{equation}
	\mathcal{H}(\mathbf{k}) = 
	\begin{bmatrix}
		0 & f_x^- & 0 & f_y^- \\
		f_x^+ & 0 & f_y^- & 0 \\
		0 & f_y^+ & 0 & f_x^- \\
		f_y^+ & 0 & f_x^+ & 0
	\end{bmatrix},
	\label{H_k_compact}
\end{equation}
with $f_{x(y)}^\pm = w_{x(y)} + v_{x(y)} e^{\pm i k_{x(y)}}$. As illustrated in Fig.~\ref{fig2}, the bulk spectrum exhibits a gap centered at $E=0$, which closes at the topological phase transition point $\beta=1$. Furthermore, $H_B$ possesses chiral symmetry $\{\Gamma, H_B\} = 0$, which topologically protects the zero-energy boundary modes against symmetry-preserving perturbations~\cite{Ryu_2010,Asb_th_2016,Li2022Topological}.

To understand the boundary physics, we examine edge states by applying open boundary conditions (OBC) along one direction while keeping PBC along the other. The corresponding semi-infinite Hamiltonians in two directions are derived respectively as~\cite{Kim_2018}
\begin{align}
	&H_{k_x}\! = \!\sum_{j, k_x} \bigg[ w_x \left( b_{j}^{\dagger}a_j + c_{j}^{\dagger}d_j \right)\! +\! w_y \left( c_{j}^{\dagger}b_j + d_{j}^{\dagger}a_j \right) \notag \\
	& \!+\! v_x e^{-ik_x} \left( a_{j}^{\dagger}b_j + d_{j}^{\dagger}c_j \right)\! +\! v_y \left( b_{j+1}^{\dagger}c_j + a_{j+1}^{\dagger}d_j \right) \!+\! \text{H.c.} \bigg], \notag \\ 
	&H_{k_y} \!=\! \sum_{i, k_y} \bigg[ w_x \left( b_{i}^{\dagger}a_i \! +\! c_{i}^{\dagger}d_i \right) + w_y \left( c_{i}^{\dagger}b_i + d_{i}^{\dagger}a_i \right) \notag \\
	& \!+\! v_x \left( a_{i+1}^{\dagger}b_i + d_{i+1}^{\dagger}c_i \right)\! +\! v_y e^{-ik_y} \left( b_{i}^{\dagger}c_i + a_{i}^{\dagger}d_i \right) \!+\! \text{H.c.} \bigg].
\end{align}

\begin{figure}
	\centering
	\includegraphics[width=0.98\linewidth]{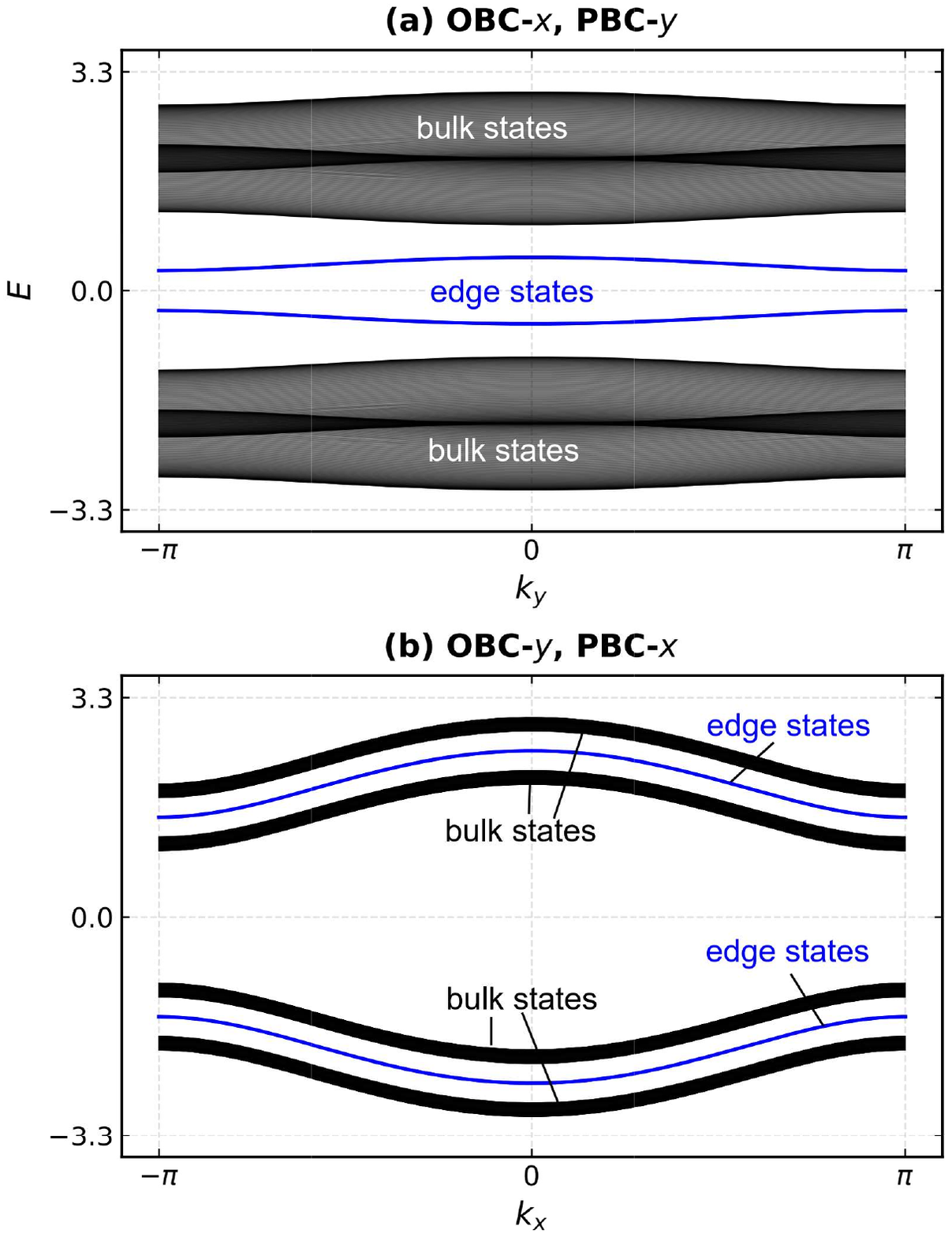}
	\caption{Energy spectrum under different boundary conditions. (a) OBC along x-direction (PBC along y): Two isolated bands appear within the bulk gap. (b) OBC along y-direction (PBC along x): Edge states merge into the bulk continuum. Blue lines highlight the edge states. Parameters are the same as in Fig.~\ref{fig2}.}
	\label{fig3}
\end{figure}

We find that the edge states exhibit strong anisotropy. For OBC along the $x$-direction, the spectrum displays two isolated bands within the bulk gap, as shown in Fig.~\ref{fig3}(a), indicating the presence of topologically protected edge states propagating along the $y$-direction. In contrast, for OBC along the $y$-direction, the edge states become dispersive and merge into the bulk continuum due to their large bandwidth exceeding the bulk gap, as shown in Fig.~\ref{fig3}(b). 

\begin{figure*}
	\centering
	\includegraphics[width=0.8\linewidth]{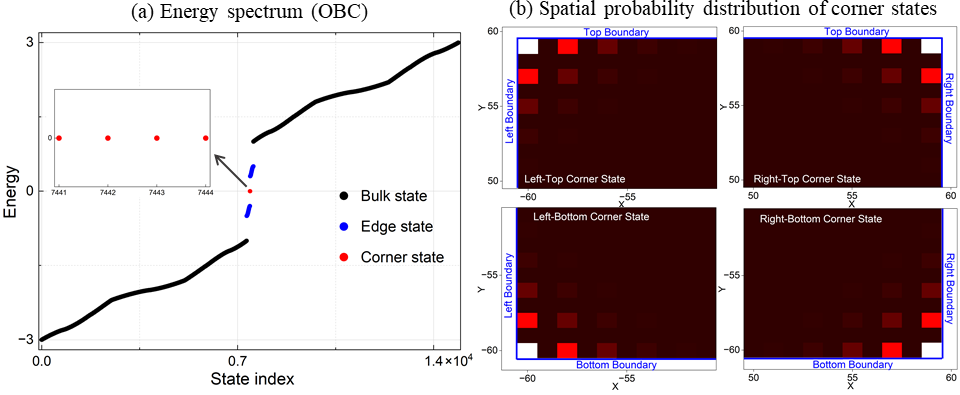}
	\caption{(a) Energy spectrum of the 2D SSH model for a $120 \times 120$ lattice with OBC in both directions. Four degenerate zero-energy states (marked by red arrows) appear at $E=0$. (b) Spatial probability distribution of four zero-energy corner states, localized at the four corners of the lattice. Parameters are identical to those in Fig.~\ref{fig2}.}
	\label{fig4}
\end{figure*}

We now consider a finite-sized system with OBC in both directions. Based on the Hamiltonian described in Eq.~(\ref{H_B}), the corresponding energy spectrum is presented in Fig.~\ref{fig4}(a). In addition to the bulk and edge bands, the system exhibits four degenerate zero-energy states at $E=0$. Figure~\ref{fig4}(b) displays the spatial probability distribution of these states, confirming that they are localized at the four corners of the lattice.

Due to chiral symmetry, the zero-energy corner states exhibit sublattice polarization in the thermodynamic limit. Taking the bottom-left corner state as an example, the wavefunction is confined to sublattice $a$, i.e., $|\psi_{\text{lb}}\rangle = \sum_{m,n} \psi_{m,n} |a_{m,n}\rangle$. Projecting the Schrödinger equation $H_B|\psi_{\text{lb}}\rangle = 0$ onto the neighboring sublattices $b$ and $d$ yields ~\cite{PhysRevResearch.7.023079,PhysRevLett.120.057001,PhysRevB.99.045441}
\begin{align}
	\langle b_{m,n}|H_B|\psi_{\text{lb}}\rangle &= w_x \psi_{m,n} + v_x \psi_{m+1,n} = 0, \notag\\
	\langle d_{m,n}|H_B|\psi_{\text{lb}}\rangle &= w_y \psi_{m,n} + v_y \psi_{m,n+1} = 0.
\end{align}
Hence $\psi_{m+1,n} = \lambda_x \psi_{m,n}$ and $\psi_{m,n+1} = \lambda_y \psi_{m,n}$ with decay factors $\lambda_x = -w_x/v_x$ and $\lambda_y = -w_y/v_y$. The normalized analytical wavefunction is
\begin{equation}
	\psi_{m,n} = \sqrt{(1-|\lambda_x|^2)(1-|\lambda_y|^2)}\, \lambda_x^{m-1} \lambda_y^{n-1},
	\label{eq:corner_wf}
\end{equation}
which exponentially localizes at the corner for $|\lambda_{x,y}|<1$. For a finite lattice of size $L_x\times L_y$, the four corner states are given by
\begin{align}
	\psi_{m,n}^{\text{lb}} &= \mathcal{N} \, \lambda_x^{m-1} \lambda_y^{n-1}, &
	\psi_{m,n}^{\text{rb}} &= \mathcal{N} \, \lambda_x^{L_x-m} \lambda_y^{n-1}, \notag\\
	\psi_{m,n}^{\text{lt}} &= \mathcal{N} \, \lambda_x^{m-1} \lambda_y^{L_y-n}, &
	\psi_{m,n}^{\text{rt}} &= \mathcal{N} \, \lambda_x^{L_x-m} \lambda_y^{L_y-n},
	\label{corner_state}
\end{align}
where the exponents represent the lattice distance from the respective corner $(1,1)$, $(L_x,1)$, $(1,L_y)$, and $(L_x,L_y)$. The normalization constant is $$\mathcal{N}=\sqrt{(1-|\lambda_x|^2)(1-|\lambda_y|^2)}.$$

\begin{figure}
	\centering
	\includegraphics[width=0.9\linewidth]{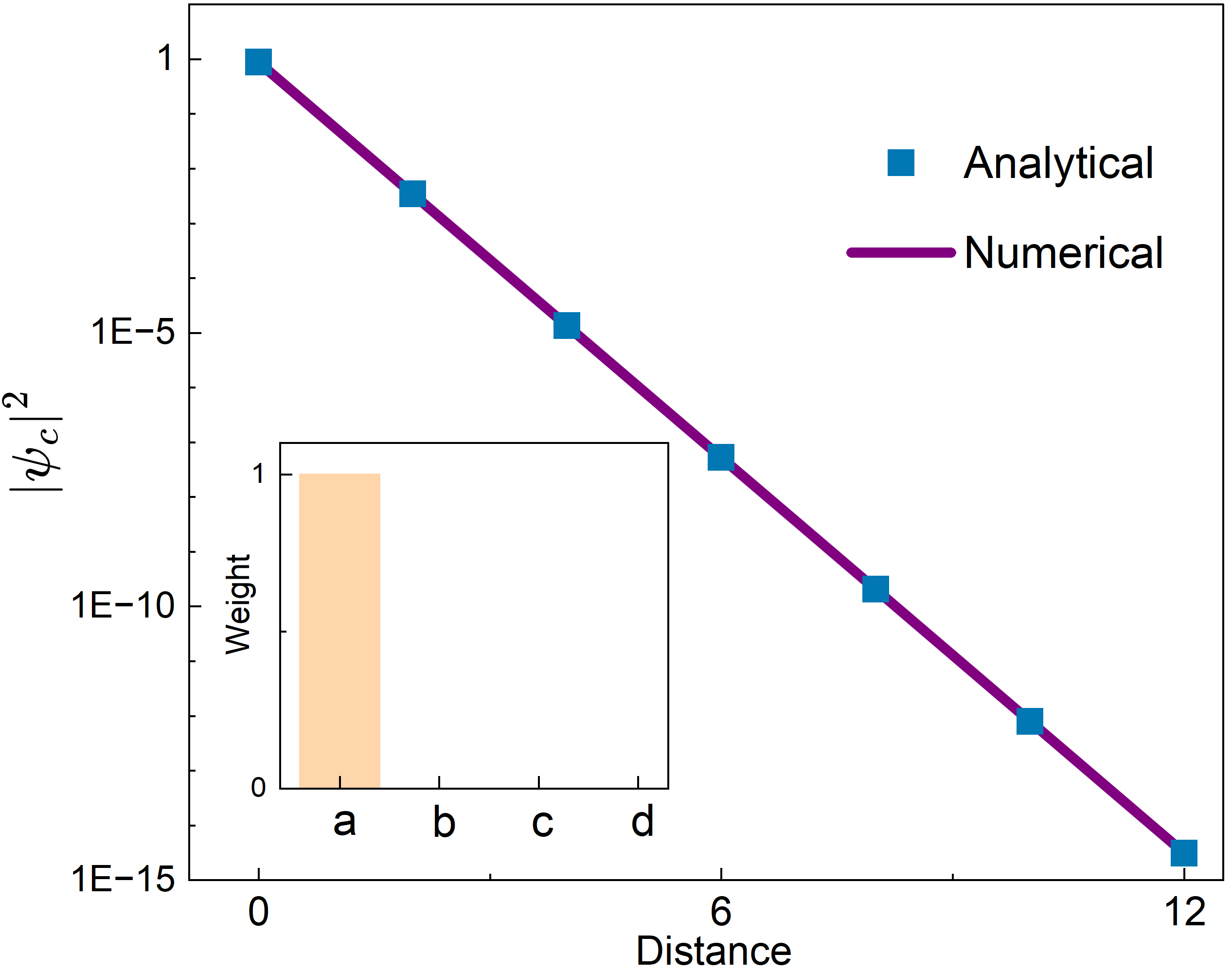}
	\caption{Numerical (symbols) and analytical (curve) corner state wavefunctions. Inset: Sublattice weight distribution showing the corner state predominantly occupies sublattice $a$. Parameters are same as in Fig.~\ref{fig2}.}
	\label{fig5}
\end{figure}

As shown in Fig.~\ref{fig5}, the numerical wavefunction matches the analytical exponential decay behavior, demonstrating the correctness of the derivation.

However, in finite-sized systems, particularly for small lattices ($N<50$), the exponential tails of the corner states overlap, leading to finite-size hybridization. The numerical eigenstates become delocalized superpositions over the four corners. By taking appropriate linear combinations of these degenerate eigenstates, we can reconstruct the single-corner localized states.

Moreover, although they have topological robustness, these corner states are limited to the physical boundaries of the lattice. This geometric confinement restricts their use in scalable quantum networks. To overcome this limitation, we propose to synthesize an effective topological boundary inside the bulk by coupling the SSH lattice to a giant atom via a vacancy-like dressed state (VDS) mechanism. The resulting system hosts corner states that are induced by the giant atom and located in the bulk, without relying on any physical edge.

\begin{figure}
	\centering
	\includegraphics[width=0.9\linewidth]{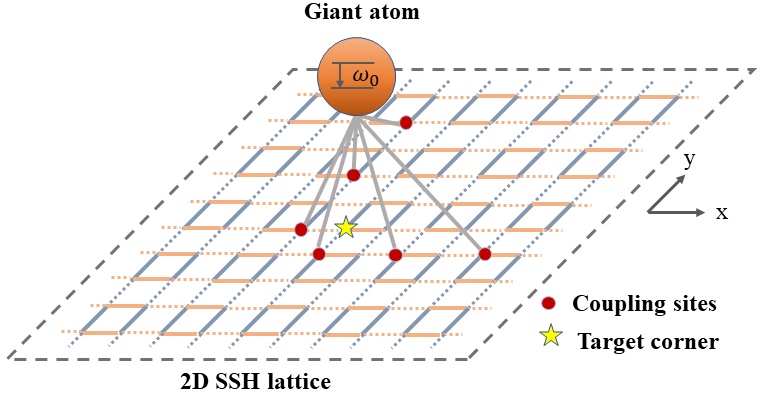}
	\caption{Schematic of a single giant atom coupled to lattice sites (red dots) arranged in an L-shaped configuration around the target corner position.}
	\label{fig6}
\end{figure}

\section{Synthesis of In-Bulk Corner States via Giant Atoms}
A VDS is an atom-photon bound state whose photonic wavefunction amplitude vanishes at the coupling sites due to destructive interference, resulting in an eigenenergy equal to the bare atomic frequency~\cite{PhysRevLett.126.063601}. Consequently, the giant atom effectively decouples from the lattice at these sites, causing the coupling points to behave as structural vacancies. By designing the spatial configuration of these coupling points, the effective vacancies can form an artificial boundary within the bulk. This artificial boundary acts as an effective open boundary condition within the bulk, introducing a corner-like feature inside the lattice. Therefore, the VDS mechanism enables the synthesis of corner states at target positions, independent of the physical edges of the system.

We demonstrate how to engineer a single corner state at the target position within the bulk of a $122 \times 122$ two-dimensional SSH lattice (to accommodate the coupling points outside the target region), as illustrated in Fig.~\ref{fig6}. To synthesize this state, we employ a giant atom with frequency $\omega_0=0$ to create an L-shaped artificial boundary. A crucial feature of this design is the preservation of chiral symmetry, which guarantees the zero-energy pinning of the engineered corner state. The 2D SSH lattice is bipartite with sublattices $A=\{a,c\}$ and $B=\{b,d\}$. In the topological non-trivial phase, the zero-energy corner states exhibit sublattice polarization, residing exclusively on one sublattice (e.g., sublattice $A$ at the inner corner). To satisfy the VDS condition without breaking chiral symmetry, we employ a complementary sublattice coupling principle. The $N_c=8$ coupling sites are chosen to reside on the complementary sublattice $B$. Because the target zero-energy mode has zero amplitude on sublattice $B$, the VDS condition $\psi_{z_i} =0$ is satisfied. Furthermore, since the corner state is exponentially localized, its wavefunction becomes negligibly small at distant sites. Thus, selecting the four nearest coupling points in each of the two orthogonal directions (extending upward and leftward from the center) is sufficient to generate an artificial boundary while keeping the required coupling strengths within the experimentally tunable range.

The total Hamiltonian of the system is written as
\begin{equation}
	H = \omega_0 |e\rangle \langle e| + H_B + \sum_{i=1}^{N_c} \left( g_i c_{z_i}^{\dagger} |g\rangle \langle e| + g_{i}^{\ast} c_{z_i} |e\rangle \langle g| \right),
	\label{eq:H_total}
\end{equation}
where $c_{z_i}$ denotes the annihilation operator at the coupling site $z_i$. We decompose the lattice Hamiltonian as $H_B = H_v + H_{Bv} + V_{v\text{-}Bv}$. Here, $H_v$ represents the on-site energy of the coupling sites, $H_{Bv}$ denotes the Hamiltonian of the lattice with these coupling sites removed, and $V_{v\text{-}Bv}$ describes the hopping between the coupling sites and their adjacent lattice neighbors.

\begin{figure*}
	\centering
	\includegraphics[width=0.9\linewidth]{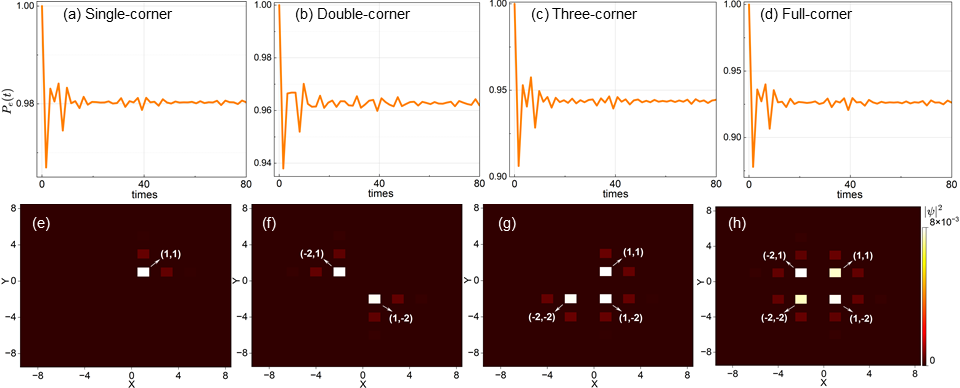}
	\caption{(a)–(d) Time evolution of the giant atom excitation probability $P_e(t)$ for single-, double-, triple-, and full-corner coupling configurations, respectively. (e)–(h) Corresponding spatial distributions of the photonic component in the lattice. Reference coupling strength is set to $g_0=0.1$. Other lattice parameters are the same as in Fig.~\ref{fig2}.}
	\label{fig7}
\end{figure*}

To derive the required coupling conditions, we define the VDS $|\Psi\rangle$ as
\begin{equation}
	|\Psi \rangle = \kappa |e\rangle \otimes |\text{vac}\rangle + |g\rangle \otimes |\psi \rangle,
	\label{eq:VDS_def}
\end{equation}
where $|\text{vac}\rangle$ denotes the photonic vacuum state, $|\psi \rangle = \sum_j \psi_j |j\rangle$ is the single-photon wavefunction in the lattice, and $\kappa$ is the probability amplitude of the atomic excited state. By definition, the VDS is an eigenstate of the total Hamiltonian with an eigenenergy equal to the bare atomic frequency $\omega_0$, satisfying
\begin{equation}
	H |\Psi\rangle = \omega_0 |\Psi\rangle.
	\label{eq:eigenvalue}
\end{equation}
Projecting Eq.~(\ref{eq:eigenvalue}) onto the atomic excited state subspace $|e\rangle \otimes |\text{vac}\rangle$, we obtain
\begin{equation}
	\kappa \omega_0 + \sum_{i=1}^{N_c} g_{i}^{\ast} \psi_{z_i} = \kappa \omega_0,
\end{equation}
where $\psi_{z_i} \equiv \langle z_i | \psi \rangle$. To ensure this equality holds for independently tunable coupling strengths $g_i$, we enforce the condition that the photonic wavefunction vanishes at each coupling site, i.e., $\psi_{z_i} = 0$. 

Geometrically, these coupling sites form an L-shaped boundary that defines an artificial corner for a $60 \times 60$ region. Inside this enclosed region, the photonic component satisfies the Schrödinger equation for a $60 \times 60$ SSH lattice with OBC. The target state is the topological corner state localized at the inner corner of the L-shape, whose analytical form is given by Eq.~(\ref{corner_state}). We denote this target wavefunction as $\psi_{\text{corner}}^{(60)}(m,n)$.

To determine the specific coupling strengths $g_i$, we further project Eq.~(\ref{eq:eigenvalue}) onto the coupling site basis $|z_i\rangle \otimes |g\rangle$, yielding
\begin{equation}
	\langle z_i | V_{v\text{-}Bv} | \psi \rangle + \kappa g_i = 0.
\end{equation}
Based on the L-shaped configuration, the interaction term $\langle z_i | V_{v\text{-}Bv} | \psi \rangle$ corresponds to the hopping from the adjacent lattice sites $p_i$ into the coupling sites $z_i$, where $p_i$ denotes the nearest-neighbor site of $z_i$ located inside the target $60 \times 60$ region. Since the wavefunction vanishes at $z_i$, this matrix element is determined by the hopping from $p_i$ to $z_i$ with strength $v_y$ (for the vertical segment, $i=1,2,3,4$) or $v_x$ (for the horizontal segment, $i=5,6,7,8$). Consequently, the required coupling strengths are given by
\begin{equation}
	g_i = 
	\begin{cases} 
		-v_y \psi_{\text{c}}^{(60)}(p_i)/\kappa, &  (i=1,2,3,4), \\
		-v_x \psi_{\text{c}}^{(60)}(p_i)/\kappa, &  (i=5,6,7,8).
	\end{cases}
	\label{eq:gi_solution}
\end{equation}
Since Eq.~(\ref{eq:gi_solution}) determines the couplings only up to a global scale, we introduce a reference coupling strength $g_0$ to fix this scale, which uniquely determines $\kappa$ and thereby all required $g_i$.

Although the coupling strengths in Eq.~(\ref{eq:gi_solution}) are derived from the eigenvalue equation at the coupling sites, the constructed state must be verified as an exact zero-energy eigenstate of the full $122\times122$ system. This is readily confirmed by a piecewise check of the Schrödinger equation. Within the $60\times60$ region, $|\psi\rangle$ is the exact corner state of the unperturbed SSH lattice and thus satisfies $H_B|\psi\rangle=0$. At the artificial boundary sites $z_i$, the eigenvalue equation $0 = \langle z_i | H_B | \psi \rangle + g_i \kappa$ is fulfilled by construction of $g_i$. In the outer region, both $\psi_j$ and $\psi_{z_i}$ vanish identically, so all hopping terms across the artificial boundary are zero. The constructed state is therefore a strict zero-energy eigenstate of the total system, confirming the exact existence of the VDS-engineered in-bulk corner state without leakage into the bulk continuum.

Moreover, this approach is generalizable, enabling the simultaneous synthesis of multiple corner states ($m \in \{1, 2, 3, 4\}$) within the bulk by increasing the number of coupling points and designing appropriate closed geometries.

Figs.~\ref{fig7}(a)--(d) show the evolution of the giant-atom excitation probability for the initial state $|e,\mathrm{vac}\rangle$. Since this state is not an eigenstate of $H$, it decomposes into the VDS and a residual component overlapping the dressed bulk bands. The residual component is rapidly emitted into the lattice, while the bound-state component remains trapped. The corresponding spatial probability distributions of the photonic component, shown in Figs.~\ref{fig7}(e)--(h), confirm that the trapped field coincides with the engineered corner states localized within the bulk. These results demonstrate that the giant-atom-mediated VDS scheme provides a flexible platform for engineering topological corner states at arbitrary positions within a photonic lattice. Note that despite the anisotropic coupling parameters ($w_x \ne w_y$), the decay factors of the zero-energy corner state are identical ($|\lambda_x|=|\lambda_y|$). Consequently, the engineered in-bulk corner state exhibits a symmetric exponential localization without any anisotropic distortion.

To assess the experimental feasibility and stability of the engineered in-bulk corner state, we examine its robustness against realistic fabrication imperfections and control noises by introducing independent random fluctuations into the multi-point coupling strengths $g_i$ and the lattice hopping parameters ($w_x, w_y, v_x, v_y$), respectively. The disordered parameters are modeled as $g_i \to g_i(1 + \delta_i^g)$ and $t_{ij} \to t_{ij}(1 + \delta_{ij}^{\text{latt}})$, where $t_{ij}$ denotes the hopping terms, with $\delta$ drawn from independent Gaussian distributions with zero means and standard deviations $\sigma_g$ and $\sigma_{\text{lattice}}$, respectively. 

For each disorder realization, we diagonalize the single-excitation Hamiltonian and identify the corner state as the eigenstate whose energy is resonant with the atomic frequency $\omega_0$ and whose photonic weight is concentrated at the engineered corner. The impact of disorder is quantified on the normalized photonic component $\psi_{\text{dis}}$ of this eigenstate. We define the fidelity with the ideal corner state, $$F = |\langle \psi_{\text{ideal}} | \psi_{\text{dis}} \rangle|^2,$$ and the inverse participation ratio (IPR) ~\cite{Sirker_2014, wegner1980inverse,Koh2024,ZHENG2024329}
\begin{equation}
	\text{IPR}(t) = \frac{\sum_{\mathbf{r}} |\psi(\mathbf{r})|^4}{\left( \sum_{\mathbf{r}} |\psi(\mathbf{r})|^2 \right)^2},
\end{equation}
which serves as a sensitive indicator of spatial localization. A high IPR confirms that the photonic state remains confined to the artificial corner, while a significant drop would indicate delocalization or scattering into the bulk continuum~\cite{rieck_2025}. All quantities are averaged over $M=50$ independent disorder configurations.

\begin{figure}[h]
	\centering
	\includegraphics[width=0.9\linewidth]{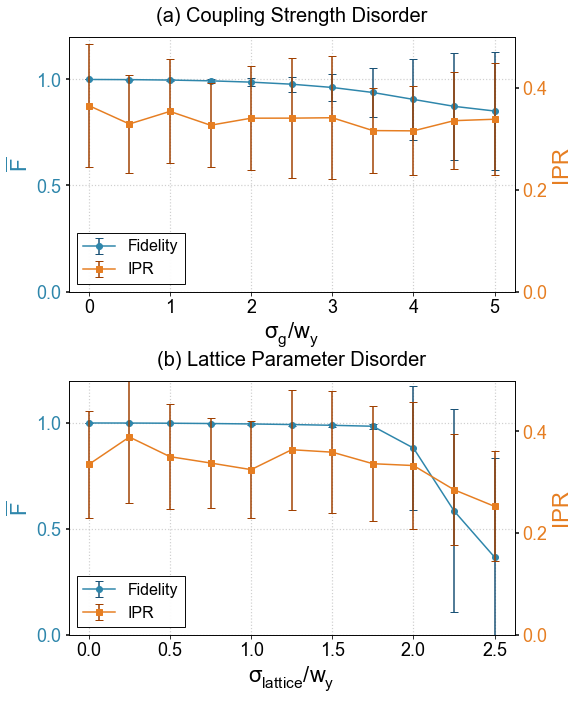}
	\caption{Ensemble-averaged dynamical fidelity $F$ (left axes, blue circles) and inverse participation ratio $\text{IPR}$ (right axes, orange squares) of the engineered corner state versus (a) the coupling disorder strength $\sigma_g$ and (b) the lattice parameter disorder $\sigma_{\rm lattice}$. The error bars denote the standard deviation over 50 independent Gaussian disorder realizations. The parameters are the same as in Fig.~\ref{fig7}.}
	\label{fig8}
\end{figure}

As shown in Fig.~\ref{fig8}, the fidelity remains high and the IPR exhibits no systematic suppression under either coupling-strength or lattice disorder, confirming that the corner state retains its spatial confinement. In summary, the VDS-based scheme enables topologically protected corner states to be engineered in the bulk of a photonic lattice, with a robustness against fabrication imperfections and control noises suitable for quantum-information applications.

\section{Dynamic Quantum Control and Multi-Atom Interactions}

Having established the synthesis of in-bulk corner states, we now extend this single-node paradigm toward advanced multi-node quantum networks. We address two fundamental requirements for scalable topological quantum information processing. The first is the versatile control over distinct topological modes, enabling the simultaneous or selective excitation of corner states and edge states within a single node. The second is the coherent, controllable coupling between spatially separated giant atoms, which is essential for distributed quantum processing. Both functionalities can be seamlessly integrated within our VDS-mediated giant atom framework through distinct multi-atom configurations.

\subsection{Multi-Channel Quantum Dynamics via Giant Superatoms }
\begin{figure}
	\centering
	\includegraphics[width=0.9\linewidth]{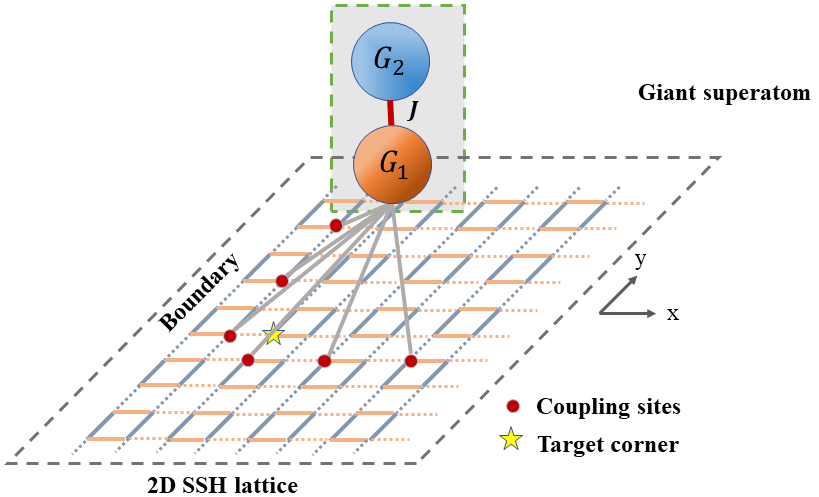}
	\caption{Schematic of a giant superatom (GSA) coupled to a 2D SSH lattice. The GSA (green dashed box) comprises two interacting atoms, $G_1$ (orange) and $G_2$ (blue), with coupling strength $J$. Atom $G_1$ is multi-point coupled to the lattice sites (red dots) arranged in an L-shaped configuration to engineer a target corner state (yellow star), while atom $G_2$ is exclusively coupled to $G_1$ and remains decoupled from the lattice.}
	\label{fig9}
\end{figure}

To achieve versatile control over distinct topological modes within a single node, we introduce a giant superatom (GSA) configuration consisting of two coupled two-level atoms, $G_1$ and $G_2$ ~\cite{crzs-k718}. As shown in Fig.~\ref{fig9}, atom $G_1$ interacts with the 2D SSH lattice via multi-point coupling at sites $\{z_i\}$ to engineer a topologically protected corner state, while atom $G_2$ exclusively couples to $G_1$ and remains decoupled from the lattice. The total Hamiltonian of this coupled system is given by
\begin{eqnarray}
	H_{\text{G}} &=& \omega_1 \sigma_1^+ \sigma^-_1 + \omega_2 \sigma_2^+ \sigma^-_2 + J \sigma_1^+ \sigma^-_2 + J^\ast \sigma_1^- \sigma^+_2 \notag \\
	&+& H_B + \sum_{i=1}^{N_c} \left( g_i c_{z_i}^{\dagger}  |g_1\rangle\langle e_1| + \text{H.c.} \right),
	\label{eq:H_GSA}
\end{eqnarray}
where $\omega_{1,2}$ are the transition frequencies, $J$ is the coherent coupling strength between the two atoms, and $g_i$ denotes the coupling between $G_1$ and the lattice sites $\{z_i\}$ arranged in an L-shape around the target corner position.

\begin{figure}
	\centering
	\includegraphics[width=0.9\linewidth]{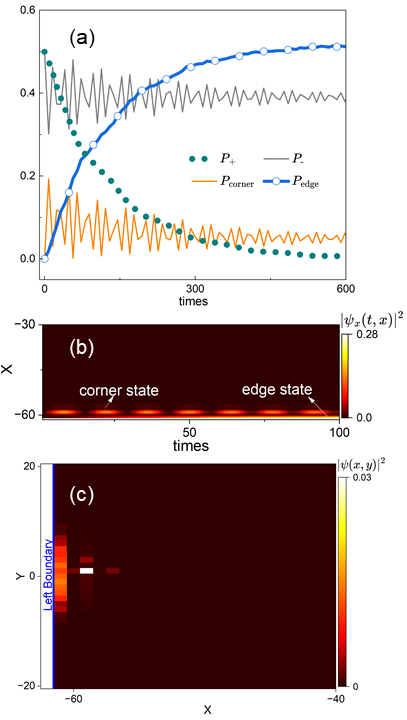}
	\caption{Multi-channel quantum dynamics of the giant superatom initialized in $|g_1 e_2\rangle$. (a) Time evolution of the dressed-state populations $P_{\pm}(t)$ together with the photonic populations of the corner mode, $P_{\rm corner}(t)$, and of the edge mode, $P_{\rm edge}(t)$. (b) Spatiotemporal photonic probability density $|\psi_x(t,x)|^2=\sum_y|\psi(t,x,y)|^2$. (c) Spatial distribution $|\psi(x,y)|^2$ at $t=40$. We set $g_1=0.5$ and $J=0.25$ to satisfy the dual-resonance condition.}
	\label{fig10}
\end{figure}

The advantage of this architecture lies in its ability to engineer a dual-resonance condition. Diagonalizing the atomic subspace yields two dressed states $|\pm\rangle = (|e_1 g_2\rangle \pm |g_1 e_2\rangle)/\sqrt{2}$ (assuming $\omega_1 = \omega_2 = \omega_0$) with eigenfrequencies $\omega_\pm = \omega_0 \pm |J|$. By tuning $\omega_0$ and $J$, we can engineer a dual-resonance condition in which $\omega_- \approx 0$ resonates with the zero-energy corner mode, while $\omega_+$ is shifted into resonance with the edge-state band.

Specifically, defining the couplings as   $g_i $ turns the $|-\rangle$ channel into a VDS in which the photonic component coincides with the vacancy-induced corner mode $\psi_c$, the energy remains pinned at $\omega_-=0$, and the field amplitude vanishes at every coupling site. The $|+\rangle$ channel hybridizes with the edge mode in the conventional resonant manner. Transforming the interaction Hamiltonian into the dressed states basis,  since the excitation operator of $G_1$ can be written as $|e_1\rangle\langle g_1| = (|+\rangle + |-\rangle)\langle g_1 g_2|/\sqrt{2}$, both dressed states couple to the exact same lattice modes $k$ with equal strength
\begin{equation}
	H_{int} = \sum_k \frac{g_k}{\sqrt{2}} c_k^\dagger |g_1 g_2\rangle (\langle +| + \langle -|) + \text{H.c.},
\end{equation}
where $g_k = \sum_i g_i \phi_k^*(z_i)$ is the effective coupling strength determined by the spatial overlap between the coupling sites and the mode wavefunction $\phi_k$. Initializing the system in $|g_1 e_2\rangle = (|+\rangle - |-\rangle)/\sqrt{2}$, the excitation is distributed equally between the two resonant channels, activating both the corner state and the edge state simultaneously without the need for complex state preparation.

As illustrated in Fig.~\ref{fig10}(a), the two channels exhibit distinct dynamics. The excitation injected into $|-\rangle$ is converted into the corner mode and remains coherent, yielding the oscillations in $P_{\rm corner}(t)$. In contrast, the edge resonance at $\omega_+=E_{\rm edge}$ hybridizes with a quasi-continuum of extended modes, so $P_{\rm edge}(t)$ rises smoothly and monotonically via irreversible loading. This phenomenon is visualized in the spatiotemporal photonic density [Fig.~\ref{fig10}(b)], where the corner state remains localized while the edge state accumulates gradually along the boundary.

Crucially, the distribution in Fig.~\ref{fig10}(c) shows that the photonic probability density simultaneously populates the localized corner position and the edge along the boundary, while diffusion into the 2D bulk remains negligible. This directly visualizes that the dual-resonance mechanism routes the excitation into both topological boundary modes at once, with the bulk bandgap providing inherent protection against decoherence.

By harnessing this mechanism, the GSA acts as a versatile quantum switch for topological modes. The excitation can be coherently trapped in the localized 0D corner state or deliberately routed into the 1D edge channels by simply tuning the interatomic coupling $J$. This multi-channel controllability within a single physical node provides a highly flexible protocol for directing quantum information flow in scalable topological networks.

\subsection{Coherent Interaction Between Giant Atoms via VDS-Mediated Corner States}

While the GSA configuration provides versatile multi-channel control, scalable quantum information processing requires coherent interactions between multiple spatially separated qubits. We now demonstrate that two independent giant atoms can interact coherently via their VDS-engineered corner states, establishing a topological quantum bus~\cite{w2w5-n3lp,photonics13020203}.

\begin{figure*}
	\centering
	\includegraphics[width=0.85\linewidth]{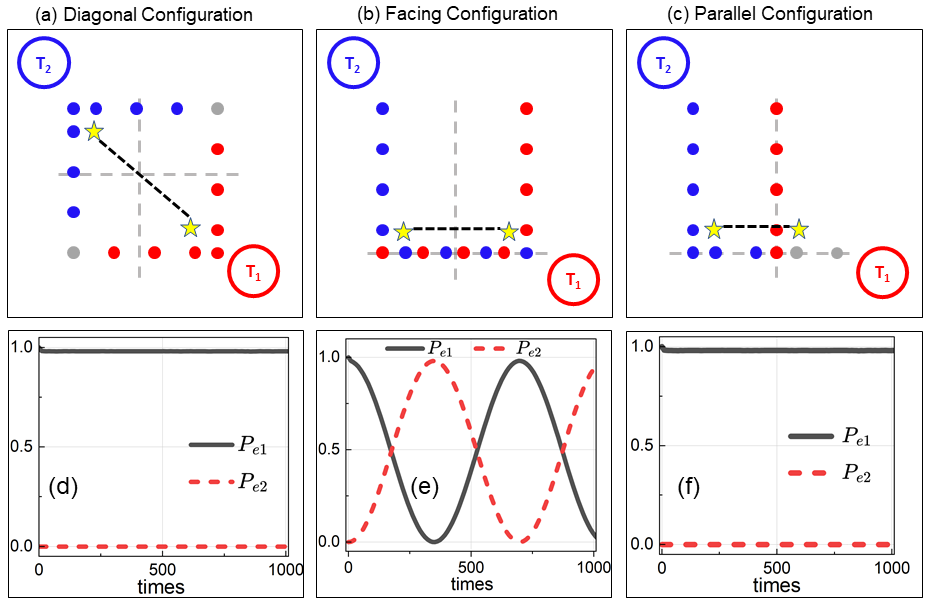}
	\caption{Coherent interaction between two giant atoms. (a)-(c) Schematic diagrams of three distinct coupling configurations for the L-shaped coupling arrays of atoms $T_1$ (red dots) and $T_2$ (blue dots). The yellow stars denote the engineered corner states. (a) Diagonal coupling configuration. (b) Facing configuration. (c) Parallel configuration. (d)–(f) Time evolution of the atomic excitation probabilities $P_{e1}$ (solid black line) and $P_{e2}$ (dashed red line) corresponding to configurations in (a)-(c), respectively. The parameters are the same as Fig.~\ref{fig7}.}
	\label{fig11}
\end{figure*}

We consider two independent giant atoms, $T_1$ and $T_2$, each coupled to its own L-shaped array of lattice sites, thereby engineering two spatially separated zero-energy corner states within the bulk. To ensure resonance with the zero-energy corner modes engineered by each atom, we set $\omega_0^{(1)} = \omega_0^{(2)} = 0$. The total Hamiltonian of the two-giant-atom system is
\begin{align}
	H &= H_B + \sum_{\eta=1}^2 \omega_0 |e_\eta \rangle\langle e_\eta| + \sum_{i=1}^{N_c} \left( g_i c_{z_i}^{\dagger} |g_1\rangle\langle e_1|
	  + \text{H.c.} \right) \notag \\&+ \sum_{j=1}^{N_c} \left( g_j' c_{z'_j}^{\dagger} |g_2\rangle\langle e_2| + \text{H.c.} \right),
	\label{eq:H_total_two_atoms}
\end{align}
Here, $N_c$ denotes the number of coupling sites for each L-shaped array, and $c_{z_i}^\dagger$ ($c_{z'_j}^\dagger$) creates a photon at the $i$-th ($j$-th) coupling site of atom $T_1$ ($T_2$).

In the weak-coupling regime ($|g_i|, |g_j'| \ll \Delta_{\mathrm{gap}}$), the lattice degrees of freedom can be adiabatically eliminated~\cite{r5zn-rqsy,
PhysRevLett.120.140404,PhysRevA.109.053720}. The Hamiltonian in the single-excitation subspace is then reduced to the effective form
\begin{equation}
	H_{\mathrm{eff}} = J_{\mathrm{eff}} \left( \sigma_+^{(1)} \sigma_-^{(2)} + \sigma_-^{(1)} \sigma_+^{(2)} \right) + \sum_{\eta=1}^2 \delta_\eta |e_\eta\rangle\langle e_\eta|,
	\label{eq:H_eff}
\end{equation}
where the coherent exchange rate $J_{\mathrm{eff}}$ is determined by the overlap of the two artificial corner-state wavefunctions, which can be expressed via the lattice Green's function as
\begin{equation}
	J_{\mathrm{eff}} = \sum_{i=1}^{N_c} \sum_{j=1}^{N_c} g_i^* g_j' G_B(z_i, z'_j; \omega=0),
	\label{eq:Jeff_analytical}
\end{equation}
where $G_B(z_i, z'_j; \omega=0) = \langle z_i | (\omega - H_B + i0^+)^{-1} | z'_j \rangle$ is the retarded Green's function of the SSH lattice evaluated at zero energy.

\begin{figure}
	\centering
	\includegraphics[width=0.95\linewidth]{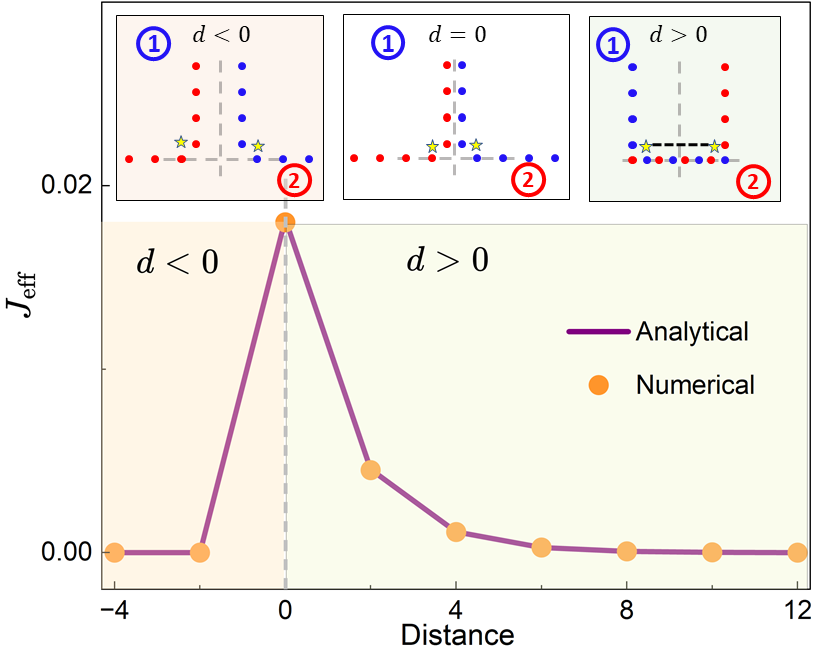}
	\caption{Distance dependence of the effective coupling strength $J_{\text{eff}}$ in the facing coupling configuration.}
	\label{fig12}
\end{figure}

To investigate the coherent interaction mechanism, we examine three distinct geometric configurations as illustrated in Fig.~\ref{fig11}. These include the diagonal configuration with point-inversion symmetry between the two L-shaped boundaries, the facing configuration with mirror symmetry, and the parallel configuration with translational symmetry. We find that the dynamics in diverse configurations are different. Coherent Rabi oscillations are observed exclusively in the facing configuration, while the interactions in both the diagonal and parallel configurations are suppressed, yielding $J_{\mathrm{eff}} \approx 0$.

Focusing on the facing configuration, we further quantify the distance dependence of the interaction strength in Fig.~\ref{fig12}. We define the separation distance $d$ between the two L-shaped boundaries, where $d < 0$ corresponds to back-to-back separated boundaries, $d = 0$ to touching boundaries, and $d > 0$ to intersecting boundaries. For $d < 0$, the coupling strength is negligibly small. Once $d \ge 0$, the interaction decreases exponentially as the separation increases. The agreement between the numerical fitting and the analytical Green's function calculation [cf. Eq.~\eqref{eq:Jeff_analytical}] confirms that the coupling is governed by the spatial overlap of the localized corner-state wavefunctions.

This phenomenon can be explained by two fundamental principles, which are the sublattice selection rule imposed by chiral symmetry, and the localization of the corner states. First, due to the bipartite nature of the SSH lattice, zero-energy corner states exhibit strict sublattice polarization. Because lattice hopping occurs exclusively between different sublattices, coherent coupling is forbidden when both corner states reside on the same sublattice. Consequently, the diagonal and parallel configurations yield $J_{\mathrm{eff}} \approx 0$, whereas the facing configuration places the corner states on opposite sublattices, satisfying the symmetry requirement for coherent interaction.

Second, even when the sublattice selection rule permits coupling, the interaction strength is governed by the spatial overlap of the wavefunctions. Each corner state decays exponentially into the region enclosed by its L-shaped boundary. For back-to-back separated boundaries ($d < 0$), the corner states decay away from each other, resulting in negligible wavefunction overlap. However, once the boundaries intersect ($d > 0$), the evanescent tails of the corner states overlap significantly within the shared region, turning on the coherent exchange. 

This geometry-dependent coupling inherently suppresses crosstalk between non-intersecting nodes. Coherent interaction is activated only on demand through intersecting boundaries with opposite sublattice polarization, providing a robust mechanism for controllable state exchange.

\section{Conclusion}

In this work, we propose a scheme to synthesize and manipulate higher-order topological corner states at arbitrary positions within the bulk of a two-dimensional Su-Schrieffer-Heeger lattice. By harnessing the VDS mechanism through engineered L-shaped multi-point couplings, a giant atom effectively induces artificial topological boundaries inside the bulk. Crucially, this approach decouples topological localization from physical geometric constraints while preserving the chiral symmetry of the lattice, ensuring that the engineered zero-energy corner states exhibit intrinsic robustness against fabrication disorder and control noise. Building upon this single-node paradigm, we extend the framework to multi-atom architectures. First, by exploiting the quantum interference within a giant superatom, we realize a versatile quantum switch that enables multi-channel control over 0D corner states and 1D edge states via dual-resonance engineering. Second, we reveal that the coherent interaction between spatially separated giant atoms is governed by the lattice's sublattice selection rule and geometric arrangement, exhibiting a strong geometry-dependent coupling with exponential distance decay. Ultimately, this work demonstrates the unique potential of giant atoms in overcoming the geometric constraints of higher-order topological insulators, providing a highly reconfigurable platform for embedding topological boundary modes within the bulk. Given the compatibility with existing photonic and superconducting circuit technologies, our scheme paves the way toward scalable topological quantum networks.
	
\section{Acknowledgment}
X.W. is supported by Natural Science Basic Research Program of Shaanxi Province (Grant No. 2026JC-YXQN-027),    Shaanxi Fundamental Science Research Project for Mathematics and Physics
(Grant No. 25JSQ026), and the National Natural Science Foundation of China (NSFC) (Grant No. 12174303).


%

\end{document}